\documentclass[preprint,showpacs,preprintnumbers,amsmath,amssymb]{revtex4}

\usepackage{graphicx}
\usepackage{dcolumn}
\usepackage{bm}

\begin{document}

\preprint{APS/123-QED}

\title{Self-Similar Solutions in the Homogeneous Isotropic Turbulence}

\author{Nicola de Divitiis}
 \altaffiliation[ ]{via Eudossiana, 18,  00184, Rome}
 \email{dedivitiis@dma.dma.uniroma1.it}
\affiliation{%
Department of Mechanics and Aeronautics\\
University "La Sapienza", Rome, Italy  
}%

\date{\today}

\begin{abstract}
We calculate the self-similar longitudinal velocity correlation function, the energy spectrum and the corresponding other properties using the results of the Lyapunov analysis of the isotropic homogeneous turbulence just presented by the author in a previous work \cite{deDivitiis2009}.
The correlation functions correspond to steady-state solutions of the evolution equation 
under the self-similarity hypothesis introduced by von K\'arm\'an.
These solutions are numerically calculated and the results adequately describe 
several properties of the isotropic turbulence.
\end{abstract}

\pacs{Valid PACS appear here}
\maketitle

\newcommand{\no}{\noindent}
\newcommand{\be}{\begin{equation}}
\newcommand{\ee}{\end{equation}}
\newcommand{\bea}{\begin{eqnarray}}
\newcommand{\eea}{\end{eqnarray}}
\newcommand{\bc}{\begin{center}}
\newcommand{\ec}{\end{center}}

\newcommand{\calr}{{\cal R}}
\newcommand{\calv}{{\cal V}}

\newcommand{\bff}{\mbox{\boldmath $f$}}
\newcommand{\bfg}{\mbox{\boldmath $g$}}
\newcommand{\bfh}{\mbox{\boldmath $h$}}
\newcommand{\bfi}{\mbox{\boldmath $i$}}
\newcommand{\bfm}{\mbox{\boldmath $m$}}
\newcommand{\bfp}{\mbox{\boldmath $p$}}
\newcommand{\bfr}{\mbox{\boldmath $r$}}
\newcommand{\bfu}{\mbox{\boldmath $u$}}
\newcommand{\bfv}{\mbox{\boldmath $v$}}
\newcommand{\bfx}{\mbox{\boldmath $x$}}
\newcommand{\bfy}{\mbox{\boldmath $y$}}
\newcommand{\bfw}{\mbox{\boldmath $w$}}
\newcommand{\bfk}{\mbox{\boldmath $\kappa$}}

\newcommand{\bfA}{\mbox{\boldmath $A$}}
\newcommand{\bfD}{\mbox{\boldmath $D$}}
\newcommand{\bfI}{\mbox{\boldmath $I$}}
\newcommand{\bfL}{\mbox{\boldmath $L$}}
\newcommand{\bfM}{\mbox{\boldmath $M$}}
\newcommand{\bfS}{\mbox{\boldmath $S$}}
\newcommand{\bfT}{\mbox{\boldmath $T$}}
\newcommand{\bfU}{\mbox{\boldmath $U$}}
\newcommand{\bfX}{\mbox{\boldmath $X$}}
\newcommand{\bfY}{\mbox{\boldmath $Y$}}
\newcommand{\bfK}{\mbox{\boldmath $K$}}

\newcommand{\bfrho}{\mbox{\boldmath $\rho$}}
\newcommand{\bfchi}{\mbox{\boldmath $\chi$}}
\newcommand{\bfphi}{\mbox{\boldmath $\phi$}}
\newcommand{\bfPhi}{\mbox{\boldmath $\Phi$}}
\newcommand{\bflambda}{\mbox{\boldmath $\lambda$}}
\newcommand{\bfxi}{\mbox{\boldmath $\xi$}}
\newcommand{\bfLambda}{\mbox{\boldmath $\Lambda$}}
\newcommand{\bfPsi}{\mbox{\boldmath $\Psi$}}
\newcommand{\bfomega}{\mbox{\boldmath $\omega$}}
\newcommand{\bfeps}{\mbox{\boldmath $\varepsilon$}}
\newcommand{\bfkappa}{\mbox{\boldmath $\kappa$}}
\newcommand{\itPsi}{\mbox{\it $\Psi$}}
\newcommand{\itPhi}{\mbox{\it $\Phi$}}
\newcommand{\bint}{\mbox{ \int{a}{b}} }
\newcommand{\ds}{\displaystyle}
\newcommand{\Sum}{\Large \sum}

\section{\bf Analysis \label{s1}}

A recent work of the author, which deals with the Lyapunov analysis of the isotropic turbulence \cite{deDivitiis2009}, suggests a mechanism for the transferring of the kinetic energy between the length scales which is based on the Landau hypothesis about the bifurcations 
of the fluid kinematic equations \cite{Landau44}.
The analysis expresses the velocity fluctuation through the Lyapunov theory and leads to the closure of the von K\'arm\'an-Howarth equation which gives the longitudinal 
velocity correlation function for two points \cite{Karman38}, i.e.
\bea
\ds \frac{\partial f}{\partial t} = 
\ds  \frac{K(r)}{u^2} +
\ds 2 \nu  \left(  \frac{\partial^2 f} {\partial r^2} +
\ds \frac{4}{r} \frac{\partial f}{\partial r}  \right) -10 \nu f \frac{\partial^2 f}{\partial r^2}(0)
\label{vk-h}  
\eea
where $K(r)$,  related to the triple velocity correlation function, realizes the closure of the Eq. (\ref{vk-h}) through the following relation \cite{deDivitiis2009} 
\bea
\ds K = u^3 \sqrt{\frac{1-f}{2}} \frac{\partial f}{\partial r}
\label{K}
\eea
and $u$ is the standard deviation of the longitudinal velocity $u_r$, which satisfies
\cite{Karman38, Batchelor53}
\bea
\ds \frac{d u^2}{d t} = 10 \nu u^2 \frac{\partial^2 f}{\partial r^2}(0) 
\eea
The skewness of $\Delta u_r$ can be expressed as \cite{Batchelor53}
\bea
\ds H_3(r) = \frac{\left\langle ( \Delta u_r )^3 \right\rangle} 
{\left\langle (\Delta u_r)^2\right\rangle^{3/2}} =
  \frac{6 k(r)}{\left( 2 (1 -f(r)  )   \right)^{3/2} }
\label{H_3_01}
\eea 
where, $k(r)$ is the longitudinal triple velocity correlation 
function, related to $K(r)$ through \cite{Batchelor53}
\bea
K(r)= u^3 \left(  \frac{\partial}{\partial r}  + \frac{4}{r}  \right) k(r)
\label{kk}
\eea
As the result, the skewness of $\partial u_r/\partial r$ is a constant which does not depend on the Reynolds number, whose value is $H_3(0) = -3/7$ \cite{deDivitiis2009}.
The other dimensionless statistical moments are consequentely determined, taking into account that the longitudinal velocity difference can be expressed as  \cite{deDivitiis2009}
\bea
\begin{array}{l@{\hspace{+0.2cm}}l}
\ds \frac {\Delta {u}_r}{\sqrt{\langle (\Delta {u}_r)^2} \rangle} =
\ds \frac{   {\xi} + \psi \left( \chi ( {\eta}^2-1 )  -  
\ds  ( {\zeta}^2-1 )  \right) }
{\sqrt{1+2  \psi^2 \left( 1+ \chi^2 \right)} } 
\end{array}
\label{fluc4}
\eea 
Equation (\ref{fluc4}) arises from statistical considerations about the Navier-Stokes equations and expresses the internal structure of the isotropic turbulence, where
 $\xi$, ${\eta}$ and $\zeta$ are independent centered random variables
which exhibit the gaussian distribution functions $p(\xi)$, $p(\eta)$ and $p(\zeta)$ whose  standard deviation is equal to the unity,
and $\psi$ is  \cite{deDivitiis2009}
\bea
\psi({\bf r}, R) =  
\sqrt{\frac{R}{15 \sqrt{15}}} \
\hat{\psi}({\bf r}, R)
\label{Rl}
\eea
The quantity  $R = u \lambda_T / \nu$ is the Taylor-scale Reynolds number, where 
$\ds \lambda_T =u/\sqrt{\langle (\partial u_r / \partial r)^2 \rangle}$ 
is the Taylor-scale, whereas the function $\hat{\psi}({\bf r}, R)$ is determined 
as $H_3(r)$ is known. 
The parameter $\chi$ is also a function of $R$ which is given by \cite{deDivitiis2009}
\bea
\frac{  8  {\psi_0}^3 \left(  1-\chi^3 \right) }
 {\left( 1+2  {\psi_0}^2 \left( 1+ \chi^2 \right) \right)^{3/2}  }
= \frac{3}{7}
\label{sk1}
\eea
with ${\psi_0} = \psi(R,0)$ and $\hat{\psi_0} = 1.075$ \cite{deDivitiis2009}.
From Eqs. (\ref{fluc4}) and (\ref{Rl}), all the absolute values of the dimensionless 
moments of $\Delta u_r$ of order greater than $3$ rise with $R$, indicating that the intermittency increases with the Reynolds number.
\\
The PDF of $\Delta u_r$ can be formally expressed with the Frobenious-Perron equation
\bea
\begin{array}{l@{\hspace{+0.3cm}}l}
F(\Delta {u'}_r) = 
\ds \int_\xi 
\int_\eta  
\int_\zeta 
p(\xi) p(\eta) p(\zeta) \
\delta \left( \Delta u_r-\Delta {u'}_r \right)   
d \xi d \eta d \zeta
\end{array}
\label{frobenious_perron}
\eea 
where $\delta$ is the Dirac delta,
whereas the spectrums $E(\kappa)$ and $T(\kappa)$ are calculated as the Fourier Transforms of $f$ and $K$ \cite{Batchelor53}, respectively, i.e.
\bea
\left[\begin{array}{c}
\ds E(\kappa) \\\\
\ds T(\kappa)
\end{array}\right]  
= 
 \frac{1}{\pi} 
 \int_0^{\infty} 
\left[\begin{array}{c}
 \ds  u^2 f(r) \\\\
 \ds K(r)
\end{array}\right]  \kappa^2 r^2 
\left( \frac{\sin \kappa r }{\kappa r} - \cos \kappa r  \right) d r 
\label{Ek}
\eea

\bigskip

\section{\bf Self-Similarity \label{s2}}

In this section the properties of the self-similar solutions of the von K\'arm\'an-Howarth
equation are studied.

Far from the initial condition, it is reasonable that the mechanism of the
cascade of energy and the effects of the viscosity act keeping $f$ and $E(\kappa)$  
similar in the time. This is the idea of self-preserving correlation function and turbulence spectrum which was originally introduced by von K\'arm\'an 
(see ref. \cite{Karman49} and reference therein).

In order to analyse this self-similarity, it is convenient
to express $f$ in terms of the dimensionless variables 
$
\ds  \hat{r} = r / \lambda_T
$
and 
$ \ds  \hat{t}  = t u /  \lambda_T$, i.e., $f = f(\hat{t}, \hat{r})$.
As the result, Eq. (\ref{vk-h}) reads as follows
\bea
\ds \frac{\partial f}{\partial \hat{t}} \ \frac{\lambda_T}{u} \frac{d}{dt} 
\left( \frac{t u}{\lambda_T} \right) -
\ds \frac{\partial f}{\partial \hat{r}} \ \frac{\hat{r}}{u} \frac{d \lambda_T}{dt}  = 
\ds  \sqrt{\frac{1-f}{2}} \frac{\partial f}{\partial \hat{r}}+
\ds  \frac{2}{R}  \left(  \frac{\partial^2 f} {\partial \hat{r}^2} +
\ds \frac{4}{\hat{r}} \frac{\partial f}{\partial \hat{r}}  \right) + \frac{10}{R} f 
\label{vk-h_adim}  
\eea
where
$
{\partial^2 f} / {\partial \hat{r}^2}(0) \equiv -1
$.
This is a non--linear partial differential equation whose coefficients 
vary in time according to the rate of kinetic energy
\bea
\ds \frac{d u^2}{d t} = - \frac{10 \nu u^2}{\lambda_T^2}
\label{rate_energy}
\eea
If the  the self--similarity is assumed, all the coefficients of Eq. (\ref{vk-h_adim})
must not vary with the time \cite{Karman38, Karman49}, thus one obtains
\bea
R = \mbox{const}
\label{coeff1}
\eea
\bea
 a_1 \equiv 
\frac{\lambda_T}{u} \frac{d}{dt} \left( \frac{t u}{\lambda_T} \right) = \mbox{const}
\label{coeff2}
\eea
\bea
a_2 \equiv \frac{1}{u} \frac{d \lambda_T}{dt} = \mbox{const}, \ \
\label{coeff3}
\eea
As the consequence of Eqs. (\ref{rate_energy}) and (\ref{coeff1}), 
 $\lambda_T$ and $u$ will depend upon the time according to
\bea
\begin{array}{l@{\hspace{-0.cm}}l}
\ds \lambda_T(t) = \lambda_T(0)\sqrt{1 +10 \nu /\lambda_T^2(0) t }, \ \ \ \ 
\ds u (t)= \frac{u(0)}{\sqrt{1+10 \nu /\lambda_T^2(0) \ t}}.
\end{array}
\eea
From these expressions, the corresponding values of $a_1$ and $a_2$ are
\bea
\begin{array}{l@{\hspace{-0.cm}}l}
\ds a_1 \equiv \frac{\lambda_T}{u} \frac{d}{dt} \left( \frac{t u}{\lambda_T} \right) =
\ds \frac{1}{1+10 \nu t / \lambda_T^2}, \\\\
\ds a_2 \equiv \frac{1}{u} \frac{d \lambda_T}{dt} = \frac{5}{R} 
\end{array}
\eea
The coefficient $a_1$ decreases with the time and for $t \rightarrow \infty$, $a_1 \rightarrow 0$,
whereas $a_2$ remains constant.
Therefore, for $t \rightarrow \infty$, one obtains the self-similar correlation function 
$f (\hat{r})$, which does not depend on the initial condition and that obeys to 
the following non--linear ordinary differential equation 
\bea
\begin{array}{l@{\hspace{-0.cm}}l}
\ds \frac{5}{R} \ \frac{d f} {d \hat{r}} \ \hat{r} +
\ds  \sqrt{\frac{1-f}{2}} \ \frac{d f} {d \hat{r}}  +
\ds \frac{2}{R}  \left(  \frac{d^2 f} {d \hat{r}^2} +
\ds \frac{4}{\hat{r}} \frac{d f}{d \hat{r}}  \right) + \frac{10}{R}  f = 0
\end{array}
\label{vk-h0}  
\eea
The first term of Eq. (\ref{vk-h0}) represents the variations in time of $f$
which does not influence the mechanism of energy cascade.
This term is negligible with respect to the second one only if $\hat{r} << R$, and is responsible 
for the asymptotic behavior of $f$ which is expressed by $\ds f \approx 1/\hat{r}^2$.
This behavior determines that all the integral scales of $f$ diverge and that $f$ does not
admit Fourier transform, thus the corresponding energy spectrum is not defined. 
According to von K\'arm\'an \cite{Karman38, Karman49}, we search the self--similar 
solutions over the whole range of $\hat{r}$, with the exception of the dimensionless distances whose order magnitude exceed $R$.
This corresponds to assume the self--similarity for all the frequencies of the energy spectrum, 
but for the lowest ones \cite{Karman38, Karman49}. 
As the result, the first term of Eq. (\ref{vk-h0}) can be neglected with respect to the
second one, and the equation for $f$ reads as follows
\bea
\begin{array}{l@{\hspace{-0.cm}}l}
\ds  \sqrt{\frac{1-f}{2}} \ \frac{d f} {d \hat{r}}  +
\ds \frac{2}{R}  \left(  \frac{d^2 f} {d \hat{r}^2} +
\ds \frac{4}{\hat{r}} \frac{d f}{d \hat{r}}  \right) + \frac{10}{R}  f = 0
\end{array}
\label{vk-h1}  
\eea
The analysis of Eq. (\ref{vk-h1}) shows that 
$
\ds f \simeq 1- {\hat{r}^2}/{2} + (10 + R)/{112} \ \hat{r}^4 
$
in the vicinity of the origin and that $\ds f -1 \approx \hat{r}^{2/3}$ when
the first term of Eq.  (\ref{vk-h1}) is about constant, 
whereas for large $\hat{r}$, $f$ exponentially decreases. Thus,  all the integral scales of $f$
are finite quantities and the energy spectrum is a definite quantity whose integral
over the Fourier space gives the turbulent kinetic energy.

\bigskip

\section{\bf Results and Discussion \label{s3}}

In this section we calculate the solutions of Eq. (\ref{vk-h1}) and study the
corresponding properties of the self--similar solutions. 
To determine $f$ and its energy spectrum, consider the following initial condition problem 
with respect to the dimensionless separation distance $\hat{r}$
\bea
\begin{array}{l@{\hspace{+0.cm}}l}
\ds  \frac{d f}{d \hat{r}} =  F \\\\
\ds \frac{d F}{d\hat{r}} = -5 f -
\left( \frac{1}{2} \sqrt{\frac{1-f}{2}} R + \frac{4}{\hat{r}} \right) F
\end{array}
\label{vk-h2}  
\eea
This ordinary differential system arises from Eq. (\ref{vk-h1}) and its
initial condition is $f(0)=1$, $F(0) = 0$. 

Several numerical solutions of Eqs. (\ref{vk-h2}) were calculated for different Taylor
scale Reynolds numbers by means of the fourth-order Runge-Kutta scheme of integration.
\begin{figure}[t]
	\centering
         \includegraphics[width=0.55\textwidth]{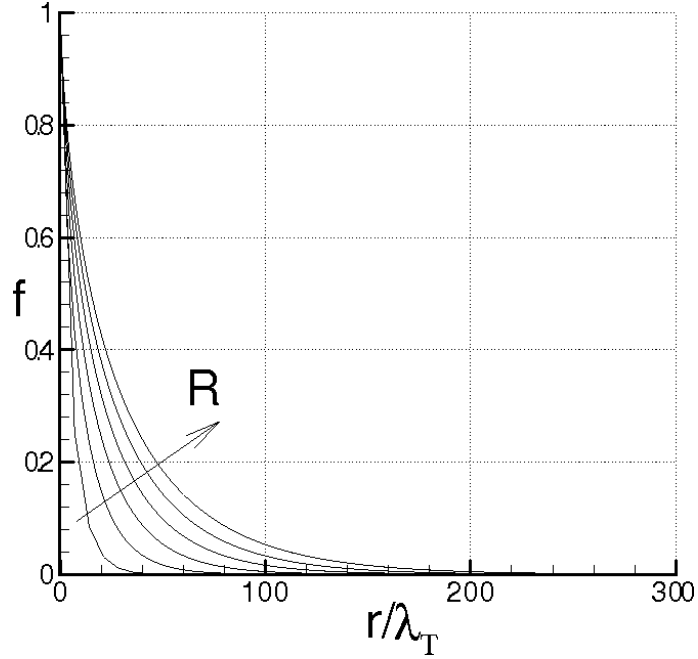}
\caption{Longitudinal correlation function for different Taylor-Scale Reynolds numbers.}
\label{figura_1}
\end{figure}
\begin{figure}[b]
	\centering
         \includegraphics[width=0.55\textwidth]{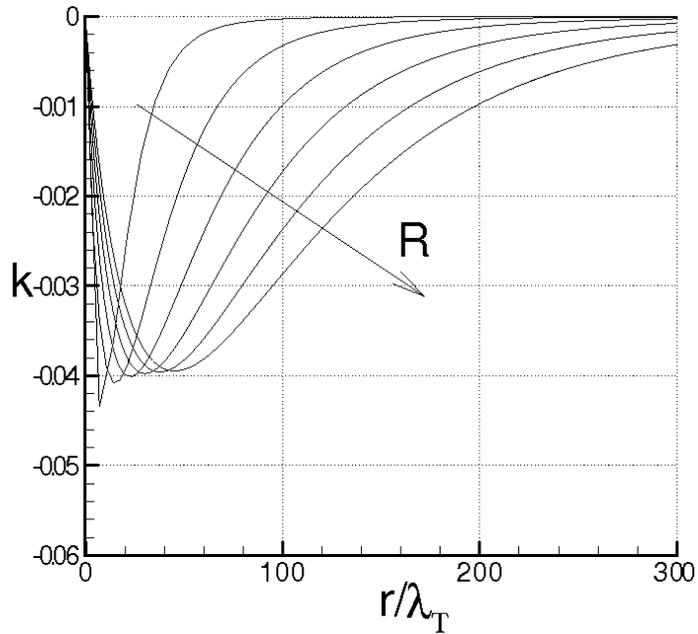}
\caption{Longitudinal triple correlation function for different Taylor-Scale Reynolds numbers.}
\label{figura_2}
\end{figure}
The cases here analyzed correspond to $R$=$100$, $200$, $300$, $400$, $500$ and $600$.
The fixed step size of the integrator scheme is selected on the basis of the asymptotic
stability condition $\Delta \hat{r} = \sqrt{2}/R$ \cite{NAG}, which also provides
a fairly accurate description of the energy spectrum at the large  wave-numbers.

Figures \ref{figura_1} and \ref{figura_2} show the numerical solutions of Eqs. (\ref{vk-h2}), where double and triple longitudinal correlation functions are represented in terms of $\hat{r}$, for the different values of $R$. Due to the mechanism of energy cascade, the tail of $f$ rises with $R$ and the maximum of $\vert k \vert$  gives the entity of this mechanism.
This value is slightly less than 0.05 and agrees quite well with the numerous data of the literature which concern the evolution of the correlation functions.
It is apparent that the spatial variations of $k$ correspond to dimensionless scales $\hat{r}$ whose size increases with $R$.

\begin{figure}[b]
	\centering
         \includegraphics[width=0.55\textwidth]{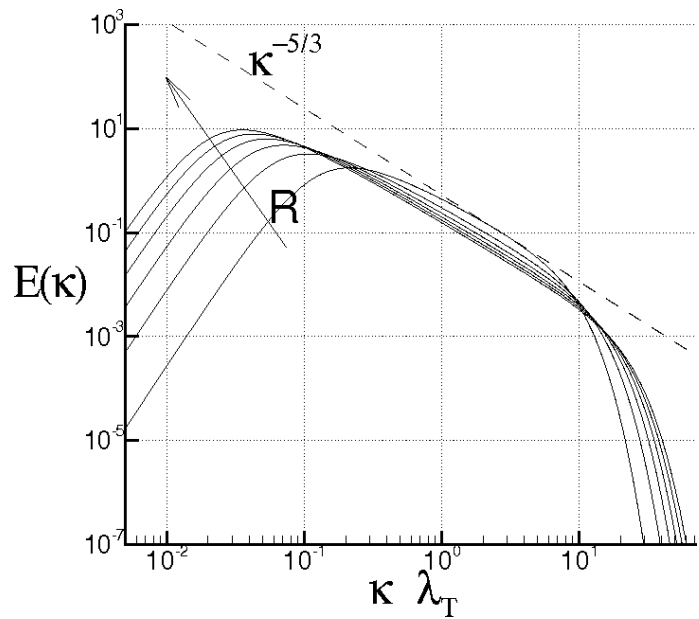}
\caption{Turbulent Energy Spectrum for different Taylor-Scale Reynolds numbers.}
\label{figura_3}
\end{figure}
\begin{figure}[t]
	\centering
         \includegraphics[width=0.55\textwidth]{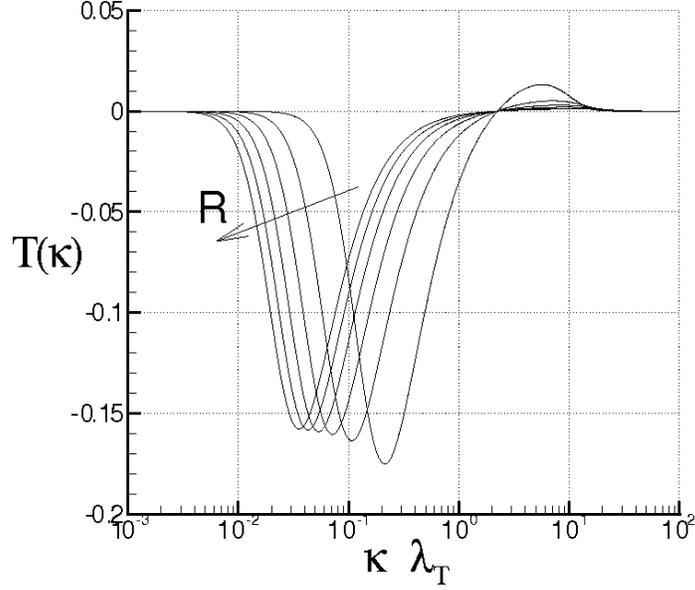}
\caption{"Transfer function $T(\kappa)$" for several Taylor-Scale Reynolds numbers.}
\label{figura_4}
\end{figure}

Figures \ref{figura_3} and \ref{figura_4} show the plots of $E(\kappa)$ and $T(\kappa)$ 
for the same Reynolds numbers.
As the consequence of the mathematical properties of $f$, the energy spectrum behaves like $E(\kappa) = O(\kappa^4)$ in proximity of the origin, and after a maximum is about parallel to the $-5/3$ Kolmogorov law (dashed line in Fig. \ref{figura_3}) in a given interval of the wave-numbers.
This interval defines the inertial range of Kolmogorov, and its size increases with $R$. 
For higher wave-numbers the energy spectrum rapidly decreases with a slope
which depends on the behavior of $f$ in proximity of the origin and thus on the Reynolds number.

Since $K$ does not modify the kinetic energy of the flow, according to Eq. (\ref{K}), 
the integral of $T(\kappa)$ over the Fourier wave-numbers results to be identically equal to zero
at all the Reynolds numbers.
\begin{figure}[b]
	\centering
         \includegraphics[width=0.55\textwidth]{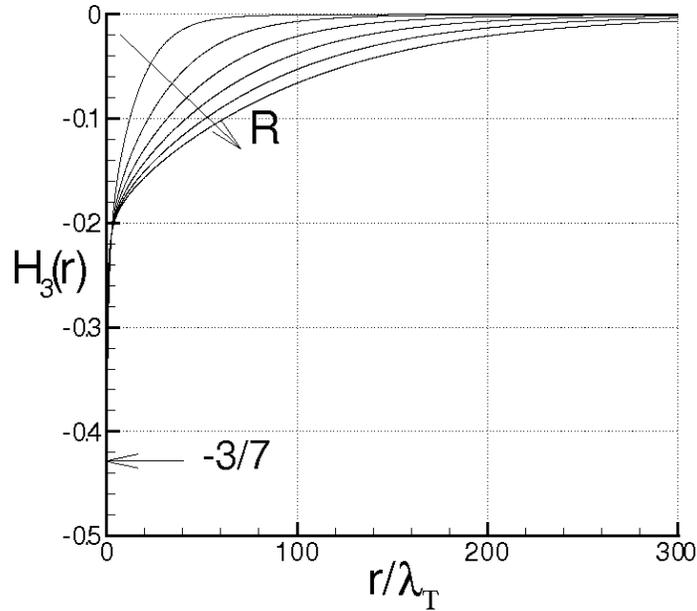}
\caption{Skewness of $\Delta u_r$ at different Taylor-Scale Reynolds numbers.}
\label{figura_5}
\end{figure}
\begin{figure}[t]
	\centering
         \includegraphics[width=0.55\textwidth]{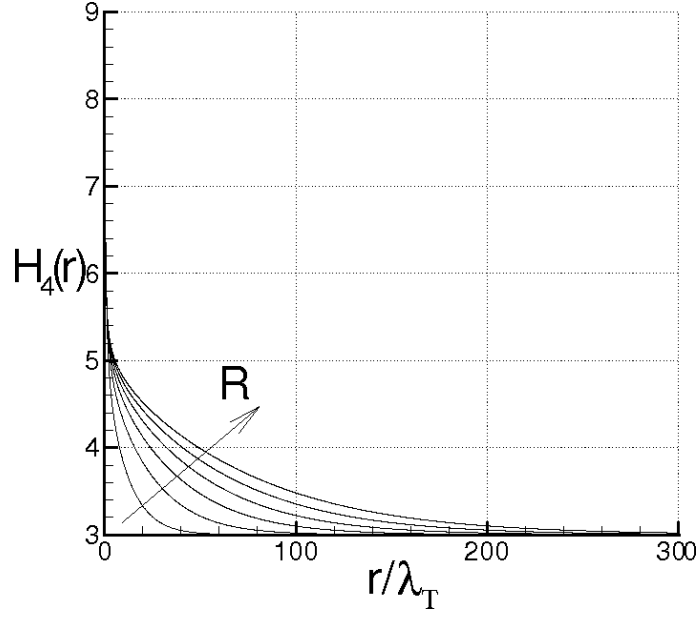}
\caption{Flatness of $\Delta u_r$ at different Taylor-Scale Reynolds numbers.}
\label{figura_6}
\end{figure}

In the Figs. \ref{figura_5} and \ref{figura_6}, skewness and flatness of $\Delta u_r$ are 
shown in terms of $\hat{r}$ for the same values of $R$.
The skewness $H_3$ is first calculated according to Eq. (\ref{H_3_01}) and 
thereafter the flatness $H_4$ has been determined using Eq. (\ref{fluc4}). 
For a given $R$, $\vert H_3 \vert$ starts from 3/7 at the origin, then decreases to small values, while $H_4 $ starts from values quite greater than 3 at $r=0$, then reaches the value of 3 (faster than $H_3$ tending to zero).
Although $H_3(0)$ does not depend upon $R$, $H_3(\hat{r})$ is a rising function
of $R$ and, in any case, the intermittency of $\Delta u_r$ increases with $R$
according to Eqs. (\ref{fluc4}) and (\ref{Rl}).

\begin{figure}[b]
	\centering
         \includegraphics[width=0.55\textwidth]{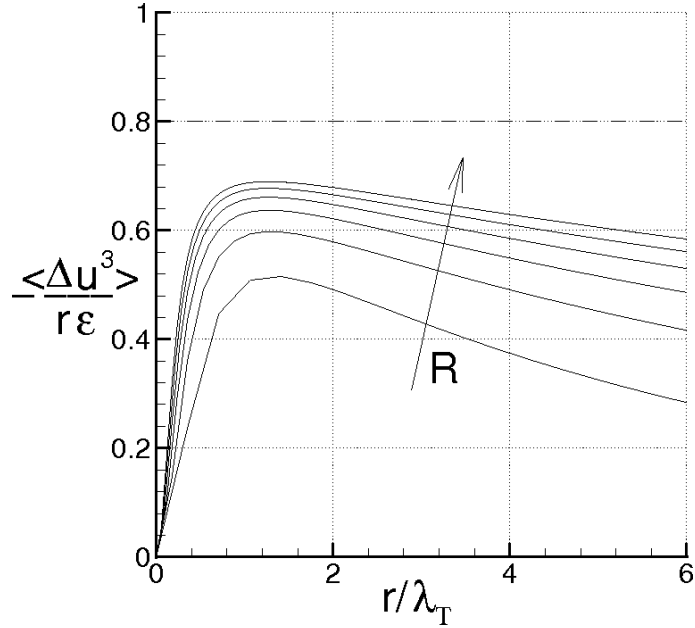}
\caption{Kolmogorov function for several Taylor-Scale Reynolds numbers.}
\label{figura_7}
\end{figure}

Next, the Kolmogorov function $Q(r)$ and Kolmogorov constant $C$, are determined using 
the previous results. 
According to the theory, the Kolmogorov function, defined as
\bea 
\ds Q(r) = - \frac{\langle (\Delta u_r)^3 \rangle} { r \varepsilon}
\label{k_f}
\eea
is constant with respect to $r$, and is equal to 4/5 as long as $r/\lambda_T = O(1)$.
As shown in Fig. \ref{figura_7}, $Q(r)$ exhibits a maximum for $\hat{r} = O(1)$ and
quite small variations for higher $\hat{r}$, as the Reynolds number increases. 
This maximum increases with $R$, and seems to tend toward the limit $4/5$ 
prescribed by the Kolmogorov theory. 

The Kolmogorov constant $C$, defined by 
$
\ds E(\kappa) \approx C {\varepsilon^{2/3} } / {\kappa^{5/3}}
$,
is here calculated as
\bea
C = \max_{\kappa \in (0, \infty)} \frac{E(\kappa) \kappa^{5/3}}{\varepsilon^{2/3}} 
\eea
where $\ds \varepsilon = - 3/2 \ d u^2/ dt$ is the rate of the energy of dissipation.
In the table \ref{table1}, the Kolmogorov constant is reported in terms of the Taylor-scale
Reynolds number.
\begin{table}[t]
  \begin{tabular}{cc} 
\hline
$R$ \       &  \ $C$ \\[2pt] 
\hline
\hline
100 \       & \ 1.8860      \\
200 \       & \ 1.9451      \\
300 \       & \ 1.9704      \\
400 \       & \ 1.9847      \\
500 \       & \ 1.9940    \\
600 \       & \ 2.0005    \\
\hline
 \end{tabular}
\caption{Kolmogorov constant for different Taylor-Scale Reynolds number.}
\label{table1}
\end{table} 
The obtained values of $C$ and $Q_{max}$ are in good agreement with the corresponding values known from the various literature.

The spatial structure of $\Delta u_r$, expressed by Eq. (\ref{fluc4}), is also studied 
with the previous results. 
\begin{figure}[t]
	\centering
         \includegraphics[width=0.55\textwidth]{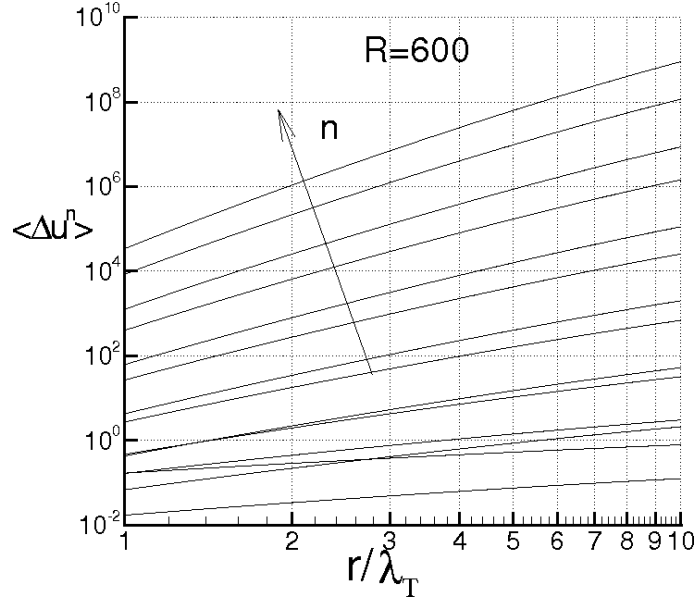}
\caption{Statistical moments of $\Delta u_r$ in terms of
the separation distance, for $R$=600.}
\label{figura_8}
\end{figure}

\begin{table}[b]
  \begin{tabular}{cccccccccccccccc} 
\hline
R & \  $\zeta(1)$ &  \ $\zeta(2)$ & \ $\zeta(3)$ & \ $\zeta(4)$ & \ $\zeta(5)$ & \ $\zeta(6)$ & \ $\zeta(7)$ & \ $\zeta(8)$ & \ $\zeta(9)$ & \ $\zeta(10)$ & \ $\zeta(11)$ & \ $\zeta(12)$ & \  $\zeta(13)$ & \ $\zeta(14)$ & \ $\zeta(15)$ \\
\hline
100 & \  0.35 & \ 0.70 & \ 1.00 & \ 1.30 & \ 1.56 & \ 1.82 & \ 2.06 & \ 2.31 & \ 2.53 & \ 2.76 & \ 2.97 & \ 3.18 & \ 3.39 & \ 3.59 & \ 3.79 \\
200 & \  0.35 & \  0.71 & \  1.00 & \  1.29 & \  1.55 & \  1.81 & \  2.05 & \  2.28  & \ 2.50 & \  2.72 & \  2.93 & \  3.14 & \  3.33 & \  3.53 & \  3.73 \\
300 & \  0.35 & \ 0.71 & \ 1.00 & \ 1.29 & \ 1.55 & \ 1.81 & \ 2.05 & \ 2.28 & \ 2.50 & \ 2.73  & \ 2.93 & \ 3.14 & \ 3.34 & \ 3.54 & \ 3.73 \\
400 & \  0.35 & \  0.71 & \  1.00 & \  1.29 & \  1.55 & \  1.81 & \  2.04 & \  2.28 & \  2.50 & \ 
 2.72 & \  2.93 & \  3.13 & \  3.33 & \  3.53 & \  3.72 \\
500 & \  0.35 & \  0.71 & \  1.00 & \  1.29 & \  1.55 & \  1.81 & \  2.04 & \  2.28 & \  2.50 & \  2.72 & \  2.93 & \  3.13 & \  3.33 & \  3.53 & \  3.73 \\
600 & \  0.35 & \  0.71 & \  1.00 & \  1.29 & \  1.55 & \  1.81 & \  2.05 & \  2.28 & \  2.51 & \ 
 2.73 & \  2.94 & \  3.15 & \  3.35 & \  3.55 & \  3.75 \\
\hline
 \end{tabular}
\caption{Scaling exponents of the longitudinal velocity difference for several Taylor-Scale Reynolds number.}
\label{table2}
\end{table} 

\begin{figure}[t]
\vspace{-5.mm}
	\centering
         \includegraphics[width=0.55\textwidth]{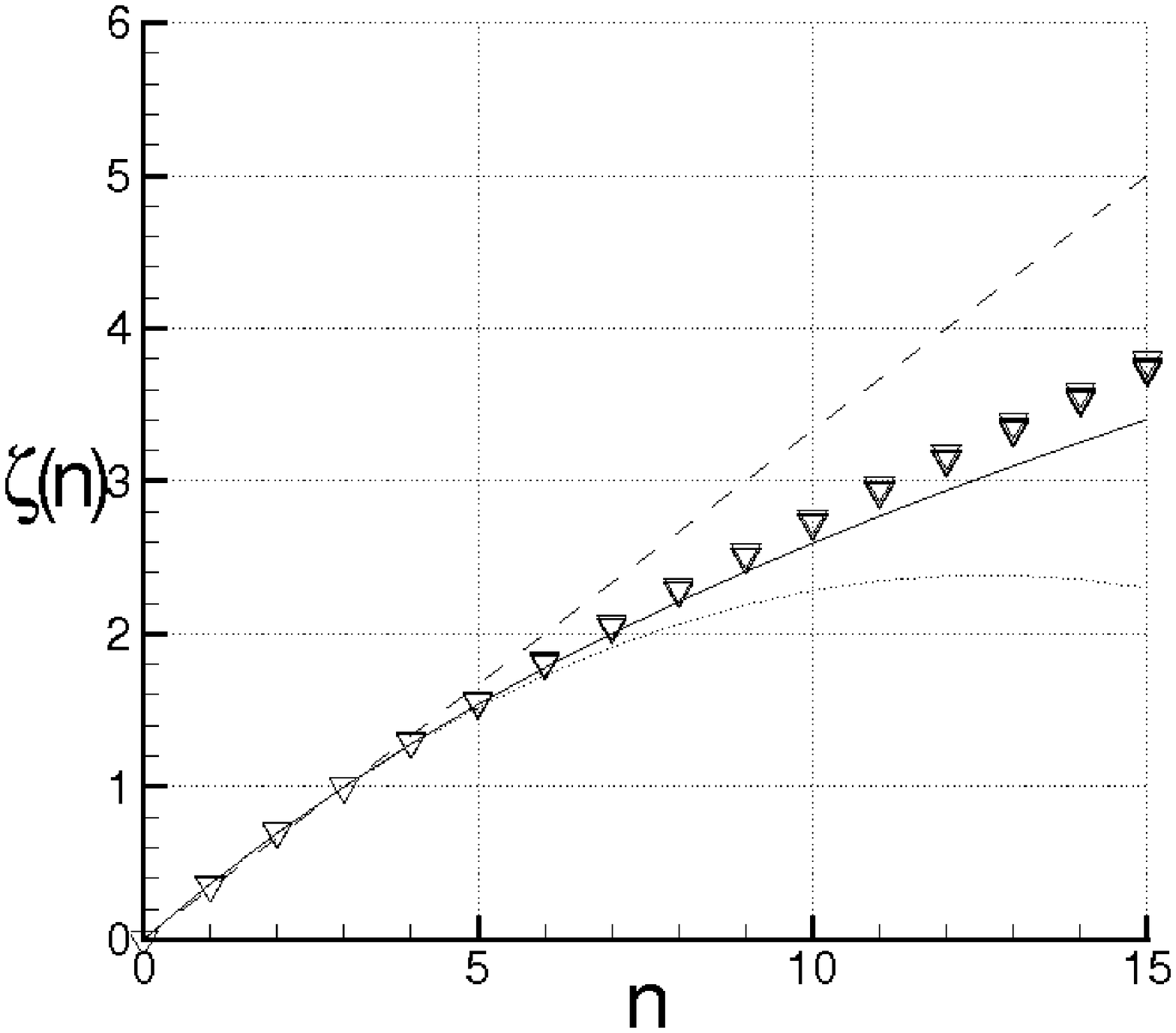}
\caption{Scaling exponents of $\Delta u_r$ for several $R$. Solid symbols are for the present data. Dashed line is for Kolmogorov K41 data \cite{Kolmogorov41}. Dotted line is for Kolmogorov K62 data \cite{Kolmogorov62}.  Continuous line is for She-Leveque data \cite{She-Leveque94}}
\label{figura_9}
\end{figure}

According to the various works \cite{Kolmogorov62, She-Leveque94, Benzi91}, 
$\Delta u_r$ behaves quite similarly to a multifractal system, where $\Delta u_r$
 obeys to a law of the kind 
$
\Delta u_r(r) \approx r^q
$
in which $q$ is a fluctuating exponent.
This implies that the statistical moments of $\Delta u_r(r)$ are expressed through 
different scaling exponents $\zeta(n)$ whose values depend on the moment order $n$, i.e.
\bea
\left\langle (\Delta u_r)^{n}(r) \right\rangle  = A r^{\zeta(n)}
\label{fractal}
\eea
\begin{figure}[t]
	\centering
         \includegraphics[width=0.55\textwidth]{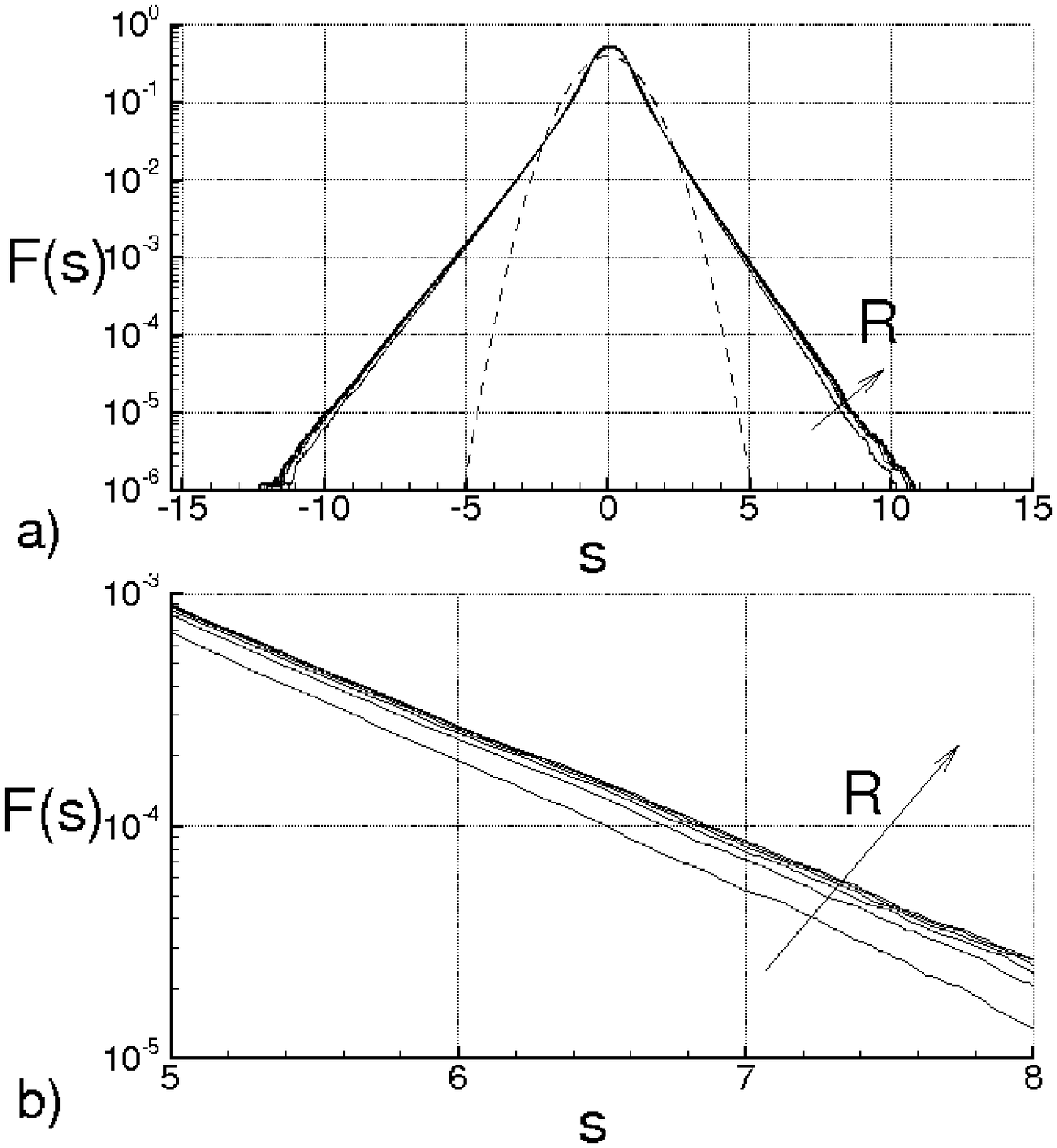}
\caption{Probability distribution functions of the longitudinal velocity derivative 
 for the different Taylor-Scale Reynolds numbers}
\label{figura_10}
\end{figure}
In order to calculate these exponents, the statistical moments of $\Delta u_r$ are first
calculated using Eqs.  (\ref{fluc4}) and (\ref{frobenious_perron}) for several separation distances.
Figure \ref{figura_8} shows the evolution of the statistical moments of $\Delta u_r$ in terms of $\hat{r}$, in the case of $R = 600$.
The scaling exponents of Eq. (\ref{fractal}) are identified through a best fitting procedure,
in the intervals ($\hat{r}_1, \hat{r}_2$), where the endpoints $\hat{r}_1$ and $\hat{r}_2$ 
have to be determined.
The calculation of $\zeta(n)$ and $A_n$ is carried out through a minimum square method which,
for each moment order, is applied to the following optimization problem
\bea
\ds J_n(\zeta(n), A_n) \hspace{-1.mm} \equiv  
\int_{\hat{r}_1}^{\hat{r}_2} 
\ds ( \langle (\Delta u_r)^n \rangle - A_n r^{\zeta(n)} )^2 dr 
 = \mbox{min}, \   n = 1, 2, ...
\eea
where $(\langle (\Delta u_r)^n)\rangle$ are calculated with Eqs. (\ref{fluc4}),
$\hat{r}_1$ is assumed to be equal to 0.1, whereas $\hat{r}_2$ is taken in such a way
that $\zeta(3)$ = 1.
The so obtained scaling exponents are shown in Table (\ref{table2}) in terms of the 
Taylor scale Reynolds number, whereas in Fig. \ref{figura_9} (solid symbols) 
these exponents are compared with those of the Kolmogorov theories K41 \cite{Kolmogorov41} (dashed line) and K62 \cite{Kolmogorov62} (dotted line), and with the exponents calculated by She-Leveque \cite{She-Leveque94} (continuous curve).
Near the origin $\zeta(n) \simeq n/3$, and in general the values of $\zeta(n)$ are in good agreement with the She-Leveque results. In particular the scaling exponents here calculated
are lightly greater than those by She-Leveque for $n >$ 8.

The PDFs of $\partial u_r /\partial {\hat{r}}$ are determined by means of Eqs. (\ref{frobenious_perron}) and (\ref{fluc4}).
Specifically, the PDF is calculated with direct simulations, where the sequences of the  variables $\xi$, $\eta$ and $\zeta$ are first determined by a gaussian random numbers generator.
The distribution function is then calculated through the statistical elaboration of the data obtained with Eq. (\ref{fluc4}).
The results are shown in Fig. \ref{figura_10}a and \ref{figura_10}b in terms of the dimensionless abscissa 
\bea
\ds s = \frac{\partial u_r /\partial {\hat{r}} } 
{ \langle \left( \partial u_r /\partial {\hat{r}} \right) ^2 \rangle^{1/2}  }
\nonumber
\eea
These distribution functions are normalized, in order that their standard 
deviations are equal to the unity. The figure represents the PDF for the several
$R$, where the dashed curve represents the gaussian distribution functions.
In particular, Fig. \ref{figura_10}b shows the enlarged region of Fig. \ref{figura_10}a,  
where  $5 < s < 8$. According to Eq. (\ref{fluc4}), the tails of PDFs change with $R$ 
in such a way that the intermittency of $\partial u_r /\partial {\hat{r}}$ rises with
the Reynolds number.

\section{\bf  Conclusions  \label{s9}}

The obtained self--similar solutions of the von K\'arm\'an-Howarth equation with
the proposed closure,  and the corresponding characteristics of the 
turbulent flow are shown to be in very good agreement with the various properties of the
turbulence from several points of view. 

In particular:

\begin{itemize}
\item 
The energy spectrum follows the Kolmogorov law in a range of wave-numbers
whose size increases with the Reynolds number.

\item
The Kolmogorov function exhibits a maximum and relatively small variations in proximity of 
$r = O(\lambda_T)$. 
This maximum value rises with the Reynolds number and seems to tend toward the limit
$4/5$, prescribed by the Kolmogorov theory.

\item 
The Kolmogorov constant moderately varies with the Reynolds number with an average
value around to 1.95 when $R$ varies from 100 to 600.

\item
The scaling exponents of the moments of velocity difference are calculated through
a best fitting procedure in an opportune range of the separation distance.
The values of these exponents are in good agreement with the results known 
from the literature.

\item
The intermittency of the longitudinal velocity difference rises with the  Reynolds number.
\end{itemize}

These results represent a further test of the analysis presented in Ref. \cite{deDivitiis2009} 
which adequately describes many of the properties of the isotropic turbulence.

\section{\bf  Acknowledgments}

This work was partially supported by the Italian Ministry for the 
Universities and Scientific and Technological Research (MIUR).


\end{document}